\documentstyle[aps]{revtex}

\def\q{{\bf q}}

\def\cs{{\chi S}}
\def\lsim{\mathrel{\rlap{\lower4pt\hbox{\hskip1pt$\sim$}}
    \raise1pt\hbox{$<$}}}
\def\gsim{\mathrel{\rlap{\lower4pt\hbox{\hskip1pt$\sim$}}
    \raise1pt\hbox{$>$}}}

\begin{document}
\input{psfig.sty}
%\draft

\title{Chiral Extrapolation, Renormalization, and the Viability of the 
 Quark Model}

\author{Adam P. Szczepaniak$^{1}$ and Eric S. Swanson$^{2}$}

\address{$^1$   Physics Department and Nuclear Theory Center \\
   Indiana University,     Bloomington, Indiana   47405-4202}
\address{$^2$ 
Department of Physics and Astronomy, University of Pittsburgh,
Pittsburgh PA 15260 and \\  Jefferson Lab, 12000 Jefferson Ave,
Newport News, VA 23606.}
%\date{\today}
\maketitle

\begin{abstract}
The relationship of the quark model to the known chiral
properties of QCD is a longstanding problem in the interpretation of
low energy QCD. In particular, how can the pion be viewed as both 
a  collective Goldstone boson quasiparticle and 
as a valence quark antiquark bound state where universal hyperfine
interactions govern spin splittings in the same way as in the heavy
quark systems. We address this issue in a simplified model which;
however, reproduces all features of QCD relevant to this problem.
A comparison of the many-body solution to our model
and the constituent quark model
demonstrates that the quark model is sufficiently flexible to describe
meson hyperfine splitting
%low energy phenomena 
provided proper renormalization conditions and
correct degrees of freedom are employed consistently.

\end{abstract}
\pacs{}
%\narrowtext

%\section{Introduction} 
One of the fundamental symmetries of the strong interaction is chiral
symmetry ($\cs$). This symmetry arises because the bare masses of the 
light $u$ and $d$ quarks  are small with respect to hadronic scales
(or, $m_{u,d} << \Lambda_{QCD}$). The nonzero quark masses imply that chiral
symmetry is explicitly broken. 
Chiral symmetry is also broken spontaneously by the QCD vacuum. This fact
is represented by the nonzero value of the quark condensate which plays the role
of an order parameter.

The phenomena associated with chiral symmetry breaking, for example 
the spectral properties of quarks in the 
dressed vacuum or the formation of Goldstone modes,     
are among the most important and most extensively studied aspects of 
strong QCD. 
As a consequence of chiral symmetry, the
axial current is (partially) conserved and the interactions of low
 momentum pions with hadrons are weak. This forms the basis of chiral 
 perturbation theory
and, more generally,  enables one to formulate an effective approach to low
energy QCD with dynamics being dominated by weakly interacting 
Goldstone bosons and nucleons~\cite{LWeinberg}. 
At high temperatures or densities 
the chiral invariance of the QCD vacuum is expected to be restored. The
thermodynamics of this phenomenon  plays 
an important role in determining signatures of the quark--gluon plasma 
in relativistic heavy ion collisions~\cite{Shur}. 
Finally, $\cs$ is also relevant for lattice gauge simulations. 
In particular, chiral symmetry predicts
a highly nonlinear dependence of a number of observables on the light quark mass. 
Since current lattice QCD simulations are limited to light quarks
with masses of the order of $200\mbox{ MeV}$ the interpretation of lattice 
results requires input from chiral symmetry-based phenomenology. 
It is clear that chiral symmetry plays a central role in determining the
structure and interactions of low energy hadrons~\cite{Thomas}.

The constituent quark model (CQM) has historically developed in parallel with chiral
theory and 
has been extensively used as a simple alternative for the study of 
hadronic phenomenology. Even though it is loosely 
connected to QCD, the quark model has had a large number of 
remarkable successes~\cite{Close}. 
The CQM is based on the idea that strong interactions lead to massive
quasiparticles (constituent quarks) and that hadronic  structure is
dominated by the interactions between valence constituent
quarks. One argues that constituent quarks are the effective
degrees of freedom arising after dynamical $\cs$
breaking due to bare quark interactions with the quark sea in the
chiral noninvariant vacuum~\cite{Finger}. In the CQM, however, 
 properties of the constituent quarks, {\it e.g.} masses and magnetic
moments, are treated as free parameters 
making the approach effectively insensitive to the underlying
chiral structure of QCD. This implies, in particular, that CQM
pions lose their nature as Goldstone bosons and are not much 
different from, say, $\rho$ mesons. The mass splitting between
the pion (a spin-0 constituent $Q{\bar Q}$ 
state) and the $\rho$  meson (a spin-1 state) is attributed to a
residual hyperfine interaction.
The hyperfine interaction is typically associated with the nonrelativistic 
reduction of single
gluon exchange between constituent quarks and yields a 
vector--pseudoscalar splitting proportional to $\alpha_s/(m_q m_{\bar q})$.

Since the notion of gluon exchange is intrinsically perturbative, it is
likely a useful concept in heavy quark systems where the large quark mass
guarantees its applicability. However its relevance to the light quark
sector is unclear. Indeed, one may expect that 
the $\pi$-$\rho$ mass splitting is
largely driven by the underlying $\cs$ rather than by the hyperfine
interaction.  
It is therefore surprising that when the splitting 
between the flavored quark-antiquark 
vector and pseudoscalar mesons is plotted against the mass of one of the 
constituent quarks (the other constituent quark mass is held fixed), it
does indeed
behave like $1/m_{con}$ typical to a hyperfine interaction
all the way down to the light quark sector~\cite{Isgur1}. This is shown by the
squares in Fig.~1. The definition of the CQM masses depends on details of the
model, here we use typical values $m_u = m_d = 330$ MeV, $m_s = 550$ MeV, $m_c = 1600$ MeV,
and $m_b = 4980$ MeV.

\begin{figure}[hbp]
\hbox to \hsize{\hss\psfig{figure=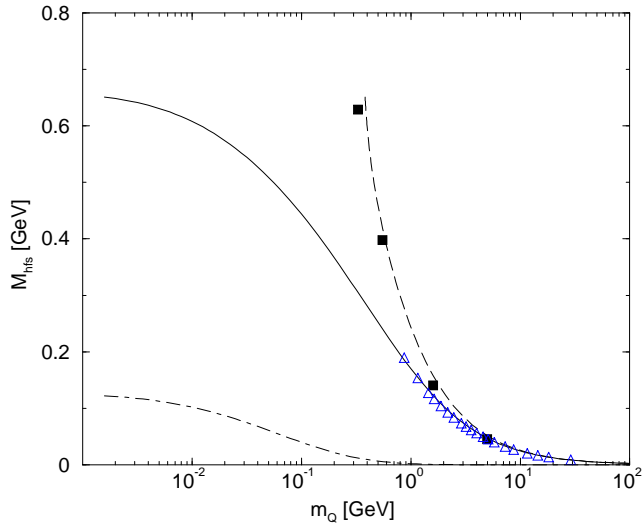,width=3.8in,angle=0}\hss}
\vspace{0.5cm}
\caption{Flavored pseudoscalar and vector meson mass 
splittings as a function  of the renormalized heavy quark mass (solid, $m_Q = m_R$) 
and constituent quark  mass (dashed, $m_Q = m_{con}$) together
with  lattice  results [12] (triangles, $m_Q = m_R$) and 
experiment [13] (squares, $m_Q = m_{con}$). The dot-dash curve represents the
hyperfine splitting obtained when the hyperfine interaction is neglected. The renormalized light quark mass is fixed at 5 MeV.}
\label{splitings}
\end{figure}

The question we address here is whether the 
apparent universality of the effective interactions of the constituent quark 
model is consistent with the dictates of chiral symmetry in the
light meson sector. To address this issue  
it is natural to work within the framework of 
Hamiltonian Coulomb gauge QCD. This enables us to map the degrees of
freedom and the global symmetries of QCD to those of the constituent quark
model and thus effectively  ``derive'' the quark model from QCD. 
The Coulomb gauge QCD Hamiltonian can be formally written as~\cite{TD,ss5}  
$H = H_{can}(\Lambda) + \delta H(\Lambda)$. 
Here the first term is the canonical Hamiltonian which contains the Coulomb 
interaction $H_{Coulomb}$ regularized at a
scale $\Lambda$. 
The cutoff $\Lambda$ is necessary since products of
field operators at the same point lead to divergences. 
Modifying the structure of the canonical Hamiltonian by the cutoff
requires counterterms, $\delta H$ which are to be adjusted so that
physical observables are $\Lambda$-independent. We note that in the chiral limit 
$\Lambda$ is the only scale in the Hamiltonian and that light 
meson masses will therefore be proportional to the cutoff. 

Even though one can in principle  choose an arbitrary value for the
cutoff,  there is typically an optimal choice, $\Lambda=\Lambda_R$ 
which depends on the  
basis and approximation scheme used to diagonalize the Hamiltonian. 
Loosely speaking, as $\Lambda$ increases beyond $\Lambda_R$, the 
strength of the  various interaction terms decrease but at the same time the
available phase space for partons to interact increases. 
On the other hand, for small cutoffs the phase
space decreases  but the remaining  Fock
sectors interact strongly, making it difficult to determine the effective
degrees of freedom. 
Since our goal is to understand the quark model from QCD, we
choose $\Lambda_R$ to match to the quark model. The quark model
scale is the scale where all QCD dynamics are given in terms of
effective interactions between valence constituent quarks (that is if
the quark model is to be a rigorous consequence of QCD). Thus, in
particular, since soft gluons are not present in the Fock space at the 
 quark model scale there are no contributions to hyperfine interactions
from soft gluon intermediate states.

We now examine the appropriate $\Lambda_R$ for heavy-heavy 
and heavy-light systems. Since the heavy-heavy case is analogous to Coulombic bound states,
it is natural to choose $\Lambda_R$ so that it scales with 
$m_R \equiv m(\Lambda=\Lambda_R)$ --  the heavy quark mass renormalized at the scale
$\Lambda_R$.  This constrains the average
momenta of heavy quarks to lie below $m_R$. Furthermore,
since interactions between heavy quarks and transverse gluons vanish
in the nonrelativistic limit, gluons and heavy quarks decouple and 
the dominant interaction between
heavy quarks is due to the nonabelian instantaneous Coulomb potential. As discussed in
detail in Ref.~\cite{Braaten}, in nonrelativistic QCD (NRQCD) the 
proper scaling of the quark-gluon interaction is obtained if 
$\Lambda_R = \alpha_R m_R$ where
$\alpha_R = \alpha(\Lambda_R) = \langle p \rangle /m_R$. Here 
$\langle p \rangle$ denotes the typical momentum of a heavy quark in
the bound state.  The scaling $\langle p \rangle \sim \alpha_R m_R$ and 
$E - 2m_R  \sim \alpha^2_R m_R$ for average momenta and 
energies is typical for a Coulombic bound
state.  

We now consider the heavy-light case. In this case the dynamics of the  
``brown  muck'' of gluons and light quarks is still 
subject to the full complexity of nonperturbative strong interactions. 
A heavy-light quark system with a single 
heavy quark is no longer Coulombic and thus the average separation between the
quarks increases causing them to probe the long range confinement potential. Quark model
phenomenology\cite{xxx} dictates that this potential be linear. Thus the relevant 
quantities are
the scales of the linear potential $b$ and the kinetic energy, which is characterized
by the reduced mass $\mu$.
Dimensional analysis then indicates that the average momenta in the bound state scale as
$\langle p \rangle = (b \mu )^{1/3}$.
Since the reduced
mass is very nearly equal to the light constituent quark mass, and this mass and
$\sqrt{b}$ are of the order of $\Lambda_{QCD}$,
we simply  use 
$\Lambda_R = \langle p \rangle \approx \Lambda_{QCD}$ for the heavy-light case. 

Proper determination of the renormalization scale is the first
step toward ``derivation'' of the quark model from QCD. The
second, is to implement dynamical chiral symmetry breaking which
will generate masses for the light bare quarks thereby creating the
constituent quarks of the CQM. The constituent Fock space then provides an
efficient basis for diagonalizing the Hamiltonian.

In this letter we do not aim at a rigorous derivation of the meson 
spectrum. Our goal is to illustrate how the two steps discussed  above
enable one to derive the constituent quark picture while maintaining  correct chiral
properties. We will therefore approximate $H$ and use a simple contact interaction 
instead of the full nonabelian Coulomb kernel.
This is sufficient as long as the interaction causes the typical momenta in the bound
state to scale as $\langle p \rangle = \alpha_R m_R$ for heavy-heavy and $\langle p \rangle = \Lambda_{QCD}$ for heavy-light mesons.

\begin{equation}
H_{Coulomb} \to H_{c}(\Lambda) = {{c_c(\Lambda)}\over {\Lambda^2}} 
 \int d{\bf x} \left[ \psi^{\dag}({\bf x}){\rm\bf T}\psi({\bf x})
\psi^{\dag}({\bf x}){\rm\bf T}\psi({\bf x}) \right]_\Lambda
\end{equation}
Here the subscript $\Lambda$ indicates that the operators are to be
point-split or smeared over a distance $\sim 1/\Lambda$\cite{ss5} and
$c_c(\Lambda)$ is the coupling which replaces $\alpha(\Lambda)$.
The counterterms $\delta H$ contain relevant, marginal, and irrelevant
operators. In the chiral limit, chiral symmetry prevents the occurrence of a 
relevant operator in the quark sector. For finite quark masses, the relevant operator
is absorbed into 
the definition of the bare quark mass. The effect of transverse
gluons eliminated by the cutoff show up through contact operators
of dimension six or greater.
There are a number of such terms; to illustrate the
effect of hyperfine
interactions we consider the dominant spin-dependent term, 
\begin{equation}
H_{h}(\Lambda) = {{c_h(\Lambda)}\over {\Lambda^2}} 
 \int d{\bf x} \left[ \psi^{\dag}({\bf x}){\rm \bf T}{\bbox{\alpha}}\psi({\bf x})
\psi^{\dag}({\bf x}){\rm \bf T}\bbox{\alpha}\psi({\bf x}) \right]_\Lambda,
\end{equation}

\noindent
where $c_h(\Lambda)$ is proportional $\alpha(\Lambda)$.
Thus the full model Hamiltonian is given by $H = \int \psi^\dagger [-i \alpha\cdot \nabla +
\beta m(\Lambda)]\psi + H_c(\Lambda) + H_h(\Lambda)$.  

As stated above, generating spontaneous chiral symmetry breaking is crucial to examining
the interplay of the hyperfine interaction and the Goldstone boson nature of the pion.
As a result, we employ the BCS ansatz to construct an approximation to the 
chirally noninvariant vacuum. 
For the above Hamiltonian this leads to the following mass gap equation which determines
the effective constituent quark mass

\begin{equation}
m_{con} = m(\Lambda) + {m_{con}\over \Lambda^2} \left( {\tilde c}_c(\Lambda)-{\tilde c}_h(\Lambda)
 \right) \int^\Lambda q^2 dq {1\over  \sqrt{m_{con}^2 + \q^2} }, \label{BCS}
\end{equation}
where ${\tilde c}_c = C_F c_c/2\pi^2$, ${\tilde c}_h = 3C_F c_h/2\pi^2$,   
and the single quasiparticle energies are given by $E(\q) = \sqrt{m_{con}^2 + \q^2}$. 
These resemble energies of the constituent quarks 
if $m_{con} \sim 300 \mbox{ MeV}$ for the light quarks. 
%Note that the renormalized
%gap equation is given by Eq. (3) with $\Lambda = \Lambda_R$.

We wish to carefully distinguish the three types of quark masses which have appeared.
They are the bare cutoff-dependent quark mass, $m(\Lambda)$ which is defined as the
parameter appearing in the mass term of the canonical Hamiltonian, the renormalized 
quark mass, $m_R$, given by $m(\Lambda_R)$, and the constituent quark mass, $m_{con}$,
which is the nonperturbative mass obtained by solving the gap equation. The cutoff 
dependence of the gap equation and $m_{con}$ are discussed later.

The RPA or Bethe-Salpeter equation for the meson bound state, $M$, is
then given by\cite{FW}

\begin{equation}
\langle M| [H,Q^{\dag}_M] | BCS\rangle = (E_M - E_{BCS} ) \langle M|
 Q^{\dag} |BCS \rangle, \label{RPA} 
\end{equation}
where $Q^{\dag}_M$ is defined in a standard way in terms of the positive and 
negative energy wave functions, $Q^{\dag}_M = \sum_{\alpha\beta} 
\left[ \psi^+_{\alpha\beta} B^{\dag}_{\alpha} D^{\dag}_{\beta} - \psi^-_{\alpha\beta} 
D_{\beta} B_{\alpha} \right]$ with $B$ and $D$ being the 
quasiparticle operators. 
In the simple approximation to $H$ used here, Eq.~(\ref{RPA}) is an algebraic
equation for the bound state masses 
$E_M = E_M(\Lambda, m(\Lambda), {\tilde c}_c(\Lambda), {\tilde c}_h(\Lambda))$.
Requiring the $E_M$ to be $\Lambda$-independent is used to determine the cutoff 
dependence of the couplings.

In the following we will concentrate on the pseudoscalar (ps) and
vector (v) open flavor mesons. 
As discussed above, in the case of unequal quark masses, the renormalization constant 
should be fixed at $\Lambda_R \approx \Lambda_{QCD}$. Here we use the numerical value
420 MeV because this corresponds to $\sqrt{b}$. The average, 
renormalized light quark mass, $m_R \equiv m(\Lambda_R)$ is set to
$5\mbox{ MeV}$,  
and the two renormalized couplings 
${\tilde c}_{R,c,h} \equiv {\tilde c}_{c,h}(\Lambda_R)$ are determined by 
fitting the $\pi$ and $\rho$ meson masses, 
$ E_{\pi,\rho} = E_{\pi,\rho}(\Lambda_R, m_R, {\tilde c}_{R,c}, {\tilde c}_{R,h})$.

Solving the gap equation yields a light constituent quark mass of 
$380 \mbox{ MeV}$.  
For a heavy-light meson, the solid line in Fig.~1 shows the dependence of the 
hyperfine mass splitting 
$M_{hfs} \equiv E_{v} - E_{ps}$ as a function of the renormalized
heavy quark mass $m_Q = m_R$ with the renormalized light quark mass fixed at 
$5\mbox{ MeV}$. 
As  $m_Q$  increases the splitting falls off as $1/m_Q$,
however, as the heavy quark mass approaches the light quark limit the
slope changes, reflecting the emergence of chiral symmetry: 
$E_{ps} \propto m_Q^{1/2}$. Our predictions reproduce the 
lattice results (triangles in Fig. 1)\cite{cjm} available for large 
quark masses.
 The lattice results are given for $M_{hfs}\cdot a $ as a
 function of $m_R \cdot a $ where $a$ is the lattice spacing. 
 Lattice calculations 
  match our results if one takes $a^{-1} = 1.45\mbox{ GeV}$
 which is very close to that of Ref.~\cite{Davis}. 
 It is nontrivial that our predictions match the lattice over a large range
of quark masses since the coupling constants were fixed in the chiral limit.
   We have fit the ansatz $M_{hfs} = a + b\sqrt{m_R +d} + c m_R$ to the data of Fig. 1.
    This ansatz has been employed in Ref.~\cite{bowler}
    to fit their lattice data and was found to be accurate for low quark masses. We similarly find it to
    be a very accurate representation of the mass splitting up to a quark mass of roughly 1 GeV.
%We note that the results shown in Fig. 1 may be used by the lattice community to extrapolate the
%hyperfine splitting to the chiral limit.

The dashed-dotted line corresponds to the mass splitting when the   
hyperfine interaction is set to zero and  
the strength of the confining term is increased so that the light 
constituent quark mass remains unchanged. It follows
that even in the chiral limit roughly $80\%$ of the $\pi-\rho$ mass 
splitting is due to the presence of the hyperfine interaction. This ratio is $92\%$ and
$99\%$ for $K^*-K$ and $D^*-D$ respectively.  We stress that the numerical 
value of this ratio depends on the 
details of the confining interaction. It is possible, for example,  to
obtain the correct $\pi-\rho$ mass splitting without a hyperfine
interaction by adjusting the strength of the confining
potential. The large $m_Q$ behavior 
of the vector-pseudoscalar mass difference would  however not be properly
described, ie., it would  decrease with $m_Q$ more rapidly than $1/m_Q$.
In this case it also found that the confining interaction
alone  leads to predictions for the quark condensate and pion decay
constant which are too small. For example, typical predictions for the condensate 
are roughly -(95 MeV)$^3$\cite{All}; which should be compared with the 
phenomenological value of -(250 MeV)$^3$.
In this model we obtain 
$\langle {\bar q} q \rangle(\Lambda_R) = -(200\mbox{ MeV})^{3}$. The improvement
is due to the hyperfine term, which has not been accounted for in previous calculations.

We now examine whether it is possible for the quark model to mimic the chiral behavior
of the hyperfine splitting shown in Fig. 1. Our hypothesis is that the freedom to 
define quark masses in the quark model is sufficient to reproduce the splitting. We address
this by plotting the hyperfine splitting as a function of the  constituent quark
mass derived from Eq. 3 (this is shown as a dashed line in Fig. 1).
The curve reproduces the observed splitting for $\rho-\pi$, $K^*-K$, $D^*-D$ and 
$B^*-B$, shown as squares in the figure.  
We conclude that it is possible for a CQM to mimic the effects of chiral symmetry
breaking. This is true because chiral symmetry breaking creates massive quasiparticles  
which may be used as effective degrees of freedom in model building. Furthermore,
the hyperfine $1/m_Q$ behavior which is valid for heavy quarks continues to be valid
for lighter constituent quarks.

We emphasize that the solid line in Fig.~1 is a prediction which may
be compared with lattice data (triangles). The only
adjustable parameters, the two renormalized potential strengths, are
fixed by the pion and rho masses when using $5\mbox{ MeV}$ for the
renormalized light quark mass.

We now consider hyperfine splittings in quarkonium systems as
a function of the quark mass.
In order to effectively mimic a Coulombic bound state for large $m_R$ and a confined 
state for small $m_R$
the renormalization scale is set to 
$\Lambda_R = {\tilde c}_{R} m_R + (m_R b)^{1/3} + b^{1/2}$.
In contrast 
with the heavy-light open flavor system, the hyperfine splitting does not vanish 
with increasing heavy quark mass $m_R$ but becomes proportional to ${\tilde c}_R^4 m_R$, as 
expected for a Coulombic system. We also considered 
the case where the renormalization scale corresponds to purely
``confined'' systems, {\it i.e.}, $\Lambda_R = (m_R b)^{1/3}
+ b^{1/2}$. This leads to 
$M_{hfs} \propto \Lambda_R^2/m_R \propto m_R^{-1/3}$, a behavior which is 
different from that expected in the constituent quark model (where 
$M_{hfs} \propto \Lambda_R^3/m^2_R \propto  m_R^{-1}$). The difference can
be traced to the presence of negative energy solutions in the RPA equation.
Elimination of the negative energy solutions leads to an effective hyperfine interaction 
which is proportional to  ${\tilde c}_{R,h}^2 \Lambda_R^2/m_R$, 
while  $H_h$ alone leads to ${\tilde c}_{R,h}\Lambda^3_R/m_R^2$. Thus it is clear that
correctly choosing the renormalization scale is crucial to establishing contact with
quark model phenomenology.

We conclude with a discussion of the renormalization procedure. 
In the simple model studied here there are three renormalization scale-dependent
quantities: $m(\Lambda)$, ${\tilde c}_c(\Lambda)$, 
and ${\tilde c}_h(\Lambda)$. We have used the  pseudoscalar and vector meson 
masses to  determine the 
(nonperturbative) renormalization scale dependence of these parameters.
The third
renormalization condition  is obtained by fixing $m_{con}$
(or alternatively $m(\Lambda_R)$ as done 
above for the light quarks) and by taking $m_{con}$ to be  
$\Lambda$-independent. For large 
$\Lambda/\Lambda_R$ Eq.~(\ref{BCS}) implies that the bare quark mass behaves as 
$m(\Lambda) \propto m_{con}^3 { \ln(\Lambda/m_{con}) } /{\Lambda^2}$.
Furthermore, at large cutoff the couplings saturate to constants. 
Thus asymptotic freedom is not recovered
in our model. This is because the contact central interaction leads to matrix elements
of order ${\tilde c}_c(\Lambda)$. In contrast, the full nonabelian Coulomb potential yields matrix 
elements of order ${\tilde c}_c(\Lambda) \ln(\Lambda)$. Asymptotic freedom is
obtained because renormalization then requires
that ${\tilde c}_c(\Lambda) \propto 1/\ln(\Lambda)$ for large $\Lambda$.
 Instead of choosing $m_{con}$ to be a physical parameter one may use 
 the condition  $d [ m(\Lambda)\langle {\bar q} q \rangle(\Lambda)]
  /d\Lambda =0$.
 This leads to a running constituent mass $m_{con} =
 m_{con}(\Lambda)$ which vanishes for large $\Lambda$. 

The use of the contact interaction rather than the full Coulomb interaction
leads to several other differences in phenomenology. For example,
the constituent quark mass becomes a function of the quasiparticle 
momentum when the Coulomb potential is employed.  Single particle energies are 
enhanced (even divergent) due to 
the confining nature of the nonabelian Coulomb potential.
Thus, in a realistic calculation
one cannot use the constituent quark energy (or constituent mass) 
to constrain 
parameters of the interaction. 
Instead one may, for example,  use the condition on the quark condensate 
given above. 
%There is also a 
%wave function renormalization which can be fixed by the pion decay
%constant, $f_\pi$. This is related to the normalization of the 
%RPA wavefunction. 
Finally, the contact interaction leads to a single bound state per
channel which certainly precludes analysis of the corresponding quark model 
excited states. 
While all of these points affect details of our computation, they do not change the
main conclusion.  

Spontaneous chiral symmetry breaking causes a 
pseudoscalar-vector meson mass splitting, but the hyperfine interaction is
required to obtain the necessary $1/m_Q$ behavior for both heavy and light 
constituent quark masses. 
The interaction between the constituent quarks
need not retain a memory of the underlying chiral symmetry and
is described well by an effective hyperfine interaction. Thus Fig. 1 represents
a direct validation of the main assumption of the naive quark model.

We have shown how quark model phenomenology may be derived from a
simple model of QCD. Elimination of high momentum components from
quark-transverse-gluon coupling leads to short range hyperfine
interactions. 
By choosing a renormalization scale which matches the quark model we neglect
contributions from long range interactions (due to the exchange of low energy
gluons) to the hyperfine
interaction  -- as is consistent with the quark model.  Then, studying 
 the quark mass dependence of the pseudoscalar-vector mass splitting 
 we are able to show that interactions between constituent quarks 
derived from QCD indeed follow that
of the naive quark model while respecting chiral symmetry.  As a result we 
have shown how the heavy
quark mass limit extrapolates to the  chiral limit and have illustrated 
the interplay between hyperfine interactions and  chiral dynamics. 
 The potential of utilizing standard many-body techniques 
 in applications to QCD in the Coulomb gauge has been discussed by 
 the authors and many others~\cite{Finger,All,cg}; however, neither the role of 
renormalization 
in constructing the Hamiltonian and building the constituent quark 
representation nor a QCD-based demonstration of the 
applicability of the effective quark model interactions to both heavy and
light quarks has been addressed in this context before.
%universality of
%the effective quark model interactions has been addressed in this 
% context before. 
 Even though the analysis presented here has used a simplified 
 central potential, the key results are independent of this choice. 
 This is because the simplified model captures all features
 ({\it i.e.}, the presence of hyperfine interactions, 
correct momentum scales and
 dynamical chiral symmetry breaking) of QCD which are relevant. A study employing the full
 nonabelian potential will be presented elsewhere.

The authors would like to acknowledge Nathan Isgur and 
Anthony Thomas for many useful discussions. 
This work was supported in part by DOE under contracts 
DE-FG02-00ER41135,  DE-AC05-84ER40150 (ES), and DE-FG02-87ER40365 (AS).


\begin{references}


\bibitem{LWeinberg} J. Gasser and H. Leutwyler, Ann. Phys. \textbf{158}, 142 (1984);
      Nucl. Phys. \textbf{B250}, 465 (1985); S. Weinberg, Phys. Lett. \textbf{B251}, 288 (1990); 
  Nucl. Phys. \textbf{B363}, 3 (1991).

\bibitem{Shur}  J. Berges and K. Rajagopal, Nucl. Phys. \textbf{B538}, 215 
(1999).

\bibitem{Thomas} D.B. Leinweber, A.W. Thomas, K. Tsushima, and S.V. Wright, 
 Phys. Rev. \textbf{D61}, 074502 (2000).

\bibitem{Close}F.E. Close, {\sl An Introduction to Quarks and Partons}  (Academic Press, New York, 1979).

\bibitem{Finger} J.R. Finger and  J.E. Mandula, Nucl. Phys. 
\textbf{B199}, 168 (1982).

\bibitem{Isgur1} This fact has been noted many times (see 
X. Song, Phys. Rev. {\bf D40}, 3655 (1989) and Ref. \cite{xxx}). Its importance for
the meaning of the quark model has been stressed by Isgur:
N. Isgur,  Nucl. Phys. \textbf{A623}, 37c (1997).

\bibitem{TD} N.H. Christ and T.D. Lee, Phys. Rev. \textbf{D22}, 939 (1980); D.G. Robertson, E.S. Swanson, 
A.P. Szczepaniak, C.R. Ji, and S.R. Cotanch, Phys. Rev. \textbf{D59}, 074019 (1999).

\bibitem{ss5} A.P. Szczepaniak and E.S. Swanson, Phys. Rev. {\bf D62}, 094027 (2000).

\bibitem{Braaten} G.T. Bodwin, E. Braaten,  and G.P. Lepage, 
 Phys. Rev. \textbf{D51}, 1125 (1995).

\bibitem{xxx} A. De Rujula, H. Georgi, and S.L. Glashow, Phys. Rev. {\bf D12}, 147 (1975);
S. Godfrey and N. Isgur, Phys. Rev. {\bf D32}, 189 (1985) and references therein.

\bibitem{FW} A.L. Fetter and J.D. Walecka, {\sl Quantum theory of many-particle systems}
(McGraw Hill, 1971).


\bibitem{cjm} J. Hein {\it et al.}, hep-ph/0003130. 

\bibitem{PDG} D.E.~Groom {\it et al.}, Eur. Phys. J. \textbf{C15}, 1 (2000).

\bibitem{Davis} C.T.H.~Davies {\it et al.},  Phys. Rev. \textbf{D56}, 2755 (1997).

\bibitem{bowler} K.C. Bowler {\it et al.}, Phys. Rev. {\bf D}, 054506 (2000). 

\bibitem{All}  A. Le Yaouanc, L. Oliver, S. Ono, O. Pene, and J.C. Raynal, 
Phys. Rev. \textbf{D31}, 137 (1985); 
F.J. Llanes Estrada and S.R. Cotanch, Phys. Rev. Lett. \textbf{84}, 1102 (2000). 

 
\bibitem{cg} 
S.L. Adler and A.C. Davis, Nucl. Phys. \textbf{B244}, 469 (1984); 
P.J.A. Bicudo and  J.E.F.T. Ribeiro,  Phys. Rev. \textbf{D42}, 
1635 (1990); Phys. Rev. \textbf{D42}, 1625 (1990);
A.P. Szczepaniak and E.S. Swanson, Phys. Rev {\bf D55}, 1578 (1997).


\end{references}
\end{document}